\documentclass{ws-ijmpd}
\usepackage[super,compress]{cite}
\usepackage{url}

\usepackage{color}

\def\comment#1{}

\begin{document}

\markboth{S.-C. Yang et al.}
{Testing dispersion of gravitational waves}

%
\catchline{}{}{}{}{}
%

\title{Testing dispersion of gravitational waves 
\\ from eccentric extreme-mass-ratio inspirals}

\author{Shu-Cheng Yang\textsuperscript{*,\dag} , Wen-Biao Han\textsuperscript{*,\dag,\S}, Shuo Xin\textsuperscript{\ddag} and Chen Zhang\textsuperscript{*,\dag}}

\address{\textsuperscript{*}Shanghai Astronomical Observatory, \\Chinese Academy of Sciences,
Shanghai, 200030, P. R. China}
\address{\textsuperscript{\dag}School of Astronomy and Space Science,\\ University of Chinese Academy of Sciences,\\
Beijing, 100049, P. R. China}

\address{\textsuperscript{\ddag}School of Physics Sciences and Engineering, \\Tongji University,
Shanghai 200092, P. R. China
\\ \textsuperscript{\S}wbhan@shao.ac.cn}

\maketitle


\begin{abstract}
In general relativity, there is no dispersion in gravitational waves, while some modified gravity theories predict dispersion phenomena in the propagation of gravitational waves. In this paper, we demonstrate that this dispersion will induce an observable deviation of waveforms if the orbits have large eccentricities. The mechanism is that the waveform modes with different frequencies will be emitted at the same time due to the existence of eccentricity. During the propagation, because of the dispersion, the arrival time of different modes will be different, then produce the deviation and dephasing of waveforms compared with general relativity. This kind of dispersion phenomena related with extreme-mass-ratio inspirals could be observed by space-borne detectors, and the constraint on the graviton mass could be improved . Moreover, we find that the dispersion effect may also be constrained by ground detectors better than the current result if a highly eccentric intermediate-mass-ratio inspirals be observed.
\end{abstract}


\ccode{PACS numbers: 04.70.Bw, 04.80.Nn, 95.10.Fh}


\section{Introduction}
After one century, the gravitational waves (GWs), which were predicted by Einstein's general relativity, have been detected by  advanced LIGO (aLIGO) and advanced Virgo (AdV) in a few events up to now\cite{abbott2016observation, abbott2016gw151226,abbott2017gw170104,abbott2017gw170608,abbott2017gw170814,abbott2017gw170817}. 6358888
ccording to general relativity (GR), GWs with different frequencies propagate at the speed of light $c$. In other words, GWs are nondispersive. In theories of quantum gravity, the graviton is the force carrier of gravity, and if GWs are nondispersive then gravitons are massless. However, in some modified gravity theories\cite{amelino2010doubly, sefiedgar2011modified, hovrava2009quantum, garattini2011modified}, GWs are dispersive. In these modified gravity theories, the rest mass of the graviton is nonzero, and the speed of the graviton depends on its frequency(energy). Consequently, the GWs would travel faster or slower than light, and their speed is frequency-related speed. 

The signals of GWs emitted from compact binaries belongs to chirp signals, of which the frequency increases over time. If GWs are dispersive and we assume that low-frequency GWs would travel slower, gravitons emitted earlier will also travel slower, leading to a frequency-dependent dephasing of GWs in GR cases.

Usually, researchers use binaries with circular orbits to test the dispersion of GWs\cite{abbott2017gw170104}, while eccentric binaries also could be employed to test the dispersion of GWs\cite{jones2004bounding}. In this physical image, one eccentric binary system will emit GWs with varied modes of frequencies at the same moment. It will lead to a more significant dephasing of the GWs in GR cases if GWs are dispersive or equivalently gravitons are massive. Generally speaking, the elliptic orbit of a comparable massive binary will be circularized at the final stage of the inspiral(where the GWs are strong enough to be detected), so that the eccentricity of the orbit could be omitted. However, for the large mass-ratio cases such as extreme-mass-ratio inspirals (EMRIs) and intermediate-mass-ratio inspirals (IMRIs), due to the formation mechanism and relatively weak radiation, it is believed that the residual eccentricity could be large at the final stage of the inspirals in most cases \cite{amaro2007intermediate}.  

In this paper, based on the physical image we discussed above, we simulate the GWs from eccentric EMRIs/IMRIs, which constrain the dispersion of GWs with higher accuracy compared with the circular cases. Because of the low frequency, this kind of GWs could be detected by space-borne interferometers such as LISA\cite{danzmann1996lisa}, Taiji\cite{hu2017the} and TianQin\cite{luo2016tianqin}. For the IMRIs with several hundred solar masses, the GWs are also locate at the sensitive band of aLIGO. Observations of such kind of eccentric IMRIs could constrain the graviton mass several times better than the current level from binary black holes.

This paper is organized as follows. Section~2 starts with the formulae of the dispersion relation and arrival time delay between different GW modes. In Sec.~3 we introduce our method for simulating gravitational waveforms and the concept of overlap and match. In Sec.~4 we present our results in details. The conclusion is given in Sec.~5.

\section{The dispersion of gravitational waves}
The distinct differences between GR and some modified gravity theories are located in the dispersion relation. In GR's theory framework, the rest mass of gravitons $m_{\rm{g}}$ must be zero like photons or other massless particles, of which the dispersion relation is $E = pc$, where $E$ and $p$ denotes the total energy and momentum of the graviton, and $c$ is the speed of light in vacuum. In some modified gravity theories $m_{\rm{g}}$ could be nonzero, and the Lorentz invariance is no longer accurate. Many methods are proposed to test Lorentz invariance, including GW observations \cite{mirshekari2012constraining} and methods that are independent of cosmological models\cite{Zhang2018Cosmological}. Since the detection of GWs, we could now test the Lorentz invariance in the dynamical sector of gravity \cite{Kostelecky2016testing, abbott2017gw170104}. To accomplish this, we consider a Lorentz-violating graviton dispersion relation\cite{mirshekari2012constraining} of the form $E^2 = p^2c^2 + m_{\rm{g}}^2c^4+\mathbb{A} p^\alpha c^\alpha$, where $\alpha$ and $\mathbb{A}$  are Lorentz-violating parameters which characterize the difference between GR and modified gravity theories. The values of $\alpha$ and $\mathbb{A}$ are different in different modified gravity theories. The speed of the graviton $v_{\rm_{g}}$ in this dispersion relation is \cite{mirshekari2012constraining}
\begin{align}
\frac{v_{\rm_{g}}^2}{c^2} = 1 - \frac{m_{\rm{g}}^2c^4}{E^2} - \mathbb{A}E^{\alpha - 2} \left(\frac{v_{\rm_{g}}}{c}\right)^{\alpha}.\label{eqn:v_g_modified}
\end{align}
The constraint on $m_{\rm{g}}$ play an important role in the dispersion of GWs. In the following, we consider the simplest situation that $\mathbb{A}$ is set to zero, in which there is no violation of Lorentz invariance. In this case, Eq.~(\ref{eqn:v_g_modified}) reduces to that of a simple massive graviton, i.e.
\begin{align}
\frac{v_{\rm{g}}^2}{c^2} = 1 - \frac{m_{\rm{g}}^2c^4}{E^2}.\label{eqn:v_g_modified2}
\end{align}

In some astrophysical events such as GW sources with electromagnetic counterparts, we could give a rough constraint on $m_{\rm{g}}$ by Eq.~(\ref{eqn:v_g_modified2}), provided that the information of the ratio of $v_{\rm_{g}}$ and $c$. The total energy of the graviton $E$ here could be acquired from the frequency $f$ of GWs by de Broglie relations. In 2017, a multi-messenger observation \cite{abbott2017gw170817, abbott2017Multi-messenger} of a binary neutron star merger found a $1.7~\rm{s}$ delay of the gamma-ray burst compared to the merger time, and the merge frequency is about $400~\rm{Hz}$. The speed difference between GWs and light in vacuum ($v_{\rm{g}} - c$) is from $-3 \times 10^{-15} $ to $+7 \times 10^{-16}$ times the speed of light\cite{abbott2017gravitational}. By using above data and Eq.~(\ref{eqn:v_g_modified2}), we get the  constraints on the rest mass of the graviton $m_{\rm{g}}\leq 1.3 \times 10^{-19}~\rm{eV}/c^2$ and the Compton wavelength of the graviton $\lambda_{\rm_{g}} > 9.7 \times 10^{9}~\rm{km}$. The accuracy of this estimation on $m_{\rm{g}}$ would be better with more such events be detected in future. In addition, the speed difference observed here is depend on the model of GW source, a model-independent measurement could be applied by observing strongly lensed GWs and the corresponding electromagnetic signals\cite{Fan2017speed}. 

Another way to constrain $m_{\rm{g}}$ focuses on the gravitational waveform of compact binaries. As mentioned above, if GWs are dispersive, then GWs with different frequencies would propagate at different velocities. During the evolution of compact binaries, the high-frequency GWs produced later would propagate faster than the low-frequency GWs produced earlier. Consequently, there would be distortion applied to the waveforms of GWs compared to that of nondispersive cases. Considering two gravitons emitted at $t_{\rm{e}}$ and $t'_{\rm{e}}$ with different frequency $f_{\rm{e}}$ and $f'_{\rm{e}}$(or with energies $E_{\rm{e}}$ and $E'_{\rm{e}}$), which will be received at corresponding arrival times $t_{\rm{a}}$ and $t'_{\rm{a}}$. Provided that during the difference of emitting time ($\Delta t_{\rm{e}} = t_{\rm{e}} - t'_{\rm{e}}$) there is little change on the scale factor $a$, then the delay of arrival times of two gravitons($\Delta t_{\rm{a}} = t_{\rm{a}} - t'_{\rm{a}}$) is \cite{mirshekari2012constraining, will1998bounding}

\begin{align}
\Delta t_{\rm{a}} = (1+Z) \left[\Delta t_{\rm{e}} + \frac{cD}{2\lambda^2_{\rm{g}}}\left(\frac{1}{f^2_{\rm{e}}}-\frac{1}{f'^2_{\rm{e}}}\right) \right]\,, \label{eqn:Delta_t_a}
\end{align}
where $Z$ is the cosmological redshift, and  
\begin{align}
D = \frac{c(1+Z)}{a_0} \int_{t_{\rm{e}}}^{t_{\rm{a}}}{a(t)dt} \,,\label{eqn:D}
\end{align}
where $a_0 = 1$ is the present value of the scale factor. For our accelerating universe \cite{riess1998observational} that is dominated by dark energy, $D$ and the luminosity distance $D_{L}$ have the forms \cite{mirshekari2012constraining, will1998bounding}
\begin{align}
D = \frac{c(1+Z)}{H_0} \int_{0}^{Z}\frac{(1 + z')^{-2}dz'}{\sqrt{\Omega_{\rm{M}}(1+z')^{3} +  \Omega_{\rm{\Lambda}}}}\label{eqn:D_equation}                          
\end{align}
and
\begin{align}
D_{L} = \frac{c(1+Z)}{H_0} \int_{0}^{Z}\frac{dz'}{\sqrt{\Omega_{\rm{M}}(1+z')^{3} +  \Omega_{\rm{\Lambda}}}},\label{eqn:D_L_equation}                          
\end{align}
 where $H_0$, $\Omega_{\rm{M}}$ and $\Omega_{\rm{\Lambda}}$ denote Hubble constant, matter density parameter today and dark energy density parameter today respectively. The radiation density parameter today $\Omega_{\rm{R}}$ is omitted here. By Eq.~(\ref{eqn:Delta_t_a}) and the observations of GW events, one could estimate the dephasing of GWs. So far aLIGO and AdV's results base on this method is  $m_{\rm{g}} \leq 7.7 \times 10^{-23}~\rm{eV}/c^2$ and $\lambda_{\rm_{g}} > 1.6 \times 10^{13}~\rm{km}$ \cite{abbott2017gw170104}. 

However, in the present work, we make a detailed analysis on dispersion by calculating the accurate waveforms from compact binaries which have large eccentricities. In this situation, the GW modes with different frequencies are emitted at the same time from this kind of eccentric sources \cite{peters1963gravitational}.  Taking $\Delta t_{\rm{e}} = 0$, Eq.~(\ref{eqn:Delta_t_a}) turns into the form
\begin{align}
\Delta t_{\rm{a}} = (1+Z) \frac{cD}{2\lambda^2_{\rm{g}}}\left(\frac{1}{f^2_{\rm{e}}}-\frac{1}{f'^2_{\rm{e}}}\right) . \label{eqn:Delta_t_a2}                          
\end{align} 
 As mentioned above, the eccentricity usually could be ignored at the detectable stage of a comparable mass-ratio binary due to the circularization of the radiation reaction. However, for the large mass-ratio cases, i.e., the EMRIs and IMRIs, the residual eccentricities could still be large at the final stage of the inspiral\cite{amaro2007intermediate}. Thus, eccentric EMRI/IMRI systems will emit GWs with varied modes of frequencies at the same moment\cite{han2017excitation}. In other words, if GWs are dispersive, gravitons that emitted at the same moment would also disperse for eccentric binary systems. 

\section{Gravitational waves from EMRIs}

The orbits of EMRIs and IMRIs could be complicated. For simplicity, in this article we only consider the situation that the eccentric orbit of the small body is set to the equatorial plane of the central body. We employ the well-known effective-one-body (EOB) theory \cite{buonanno1999effective} to describe the orbital dynamics, and specially in this work the methods we used for eccentric orbits  can be found in our previous work \cite{han2014gravitational, cao2017waveform} for EMRIs and binary black holes (SEOBNRE) respectively. The evolution of orbital parameters due to gravitational radiation for extreme mass-ratio cases are calculated from the formalisms given by Sago and Fujita \cite{sago2015calculation} (see Eq. (C6, C7) in this literature for details).

The post-Newtonian orbital evolution and waveforms for a binary system have been discussed in several papers \cite{moreno1995gravitational, barack2004lisa, pierro2001fast, yunes2009post}. In this work, we use the Teukolsky-based waveforms which are solved from the Teukolsky equations \cite{teukolsky1973perturbations}. Our method is based on frequency-domain decomposition, and has been developed in previous works \cite{han2010gravitational,han2011constructing,han2014gravitational,han2017excitation}, in which the gravitational waveform from an eccentric EMRI with total mass $M$  at distance $R$ , latitude angle $\Theta$ and azimuthal angle $\Phi$ of  could be written as (in geometrical units $G = c =1$)
\begin{align}
h_{+} - ih_{\times} = \frac{2}{R} \sum\limits_{lmk} \frac{Z^{\rm H}_{lmk}}{\omega^2_{mk}} {_{-2}S}^{a \omega }_{lmk}\left(\Theta\right){e}^{-i\phi_{mk}+i m \Phi},\label{eqn:GW_waveform}                          
\end{align}where $l,~m,~k$ are the harmonic numbers,  $\phi_{mk} \equiv \int\omega_{mk} (t) dt$,  $_{-2}S^{a \omega}_{lmk}\left(\Theta\right)$ denotes spin-weighted spheroidal harmonics which depend on the polar angles $\Theta$ of the observer's direction of sight and the direction of orbital angular momentum of the source, and $Z^{\rm H}_{lmk}$ describe the amplitude of each mode, which could be calculated by the radial component of Teukolsky equation (see Appendix A for details). In this article, we set $\Theta = 0$ (``face on'') and $\Phi=0$ without losing the generality, and $\omega_{mk}$ is 
\begin{align}
\omega_{mk} = m\Omega_{\phi} + k\Omega_{r} ,\label{eqn:graviton_omega}                          
\end{align} 
where $\Omega_{r}$ and $\Omega_{\phi}$ denote the orbital frequencies of radial and azimuthal direction respectively. For including the mass-ratio correction, we use the EOB dynamics to calculate these two fundamental frequencies, see our previous work for details \cite{han2014gravitational,han2017excitation,cao2017waveform}. In this work, the waveform with dispersion could be written as,

\begin{align}
h_{+} - ih_{\times} = \frac{2}{R} \sum\limits_{lmk} \frac{Z^{\rm H}_{lmk}}{\omega^2_{mk}} {_{-2}S}^{a \omega }_{lmk}\left(\Theta\right){e}^{-i\phi_{mk}+i m \Phi+i\omega _{mk}\Delta t_{\rm a}^{mk }},\label{eqn:GW_waveform2}                          
\end{align} where $\Delta t_{\rm a}^{mk}$ is the arrival time-delay of a specified ``voice" with harmonic number $m,~k$.  By Eq.~(\ref{eqn:Delta_t_a2}) we know $\Delta t_{\rm a}^{mk}$ is the function of two emit frequencies $ f_{\rm e}^{mk}$ and $ f_{\rm e}^{m'k'}$ of ($m, ~k$) and ($m', ~k'$) modes respectively, while $f_{\rm e}^{mk} = \omega_{mk}/ 2\pi$. In this way the dispersion effect is introduced in the waveform. Due to the orbit evolution of the binary system that caused by radiation reaction, the terms $Z^{\rm H}_{lmk}S^{\alpha \omega _{mk}}_{lmk}\left(\theta\right) $, $\omega_{mk}$ and $\Delta t_{\rm a}^{mk}$ are actually changing as the evolution of eccentricity $e$ and semi-latus rectum $p$ , and we take the waves of the largest strain\cite{fujita2009efficient} for which $l = m = 2$. In this way for a certain system state, the amplitudes of modes depend on the harmonic number $k$ \cite{han2017excitation}.

Based on above discussions, by using aLIGO and AdV's lower limit \cite{abbott2017gw170104} on $\lambda_{\rm_{g}} = 1.6 \times 10^{13}~\rm{km}$, we calculate some values of $\Delta t_{\rm{a}}$ in different orbital eccentricities for different total masses of system $M$. We use the cosmological-parameter values of WMAP(Wilkinson Microwave Anisotropy Probe) in nine-year observations\cite{hinshaw2013nine}, i.e. $H_0 = 69.3~(\rm{km}/\rm{s})/\rm{Mpc}$, $\Omega_{\rm{M}} = 0.286$ and $\Omega_{\rm{\Lambda}} = 0.714$, and in the following discussion we use this configuration to calculate $D$  and $Z$ from $D_{L}$. The maximum mode (where $k=\tilde{k}$) is specified as the reference mode, then the modes with higher ($k > \tilde{k}$) or lower ($k < \tilde{k}$) frequencies have faster or slower speed compared to the $\tilde{k}$ mode. Due to the different speed of modes, after a long distance propagation, the time delay of these modes may be obvious for observation. Some results of $\Delta t_{\rm a}$ are shown in Tables~\ref{tab:table1} and \ref{tab:table2} for $e = 0.5$ and $e = 0.7$, respectively, and the semi-latus rectum $p=12~M$(in the units with $G = c = 1$, the same as follows). Notice that the delays of arrival time $\Delta t_{\rm{a}}$ of different GW modes increase with $M$. The higher mass corresponding to lower frequency GWs, and by Eq.~(\ref{eqn:Delta_t_a2}) we know that a lower $f_{\rm{e}}$  leads to a larger $\Delta t_{\rm{a}}$. If the sources are located at $1~\rm{Gpc}$, we could see for the $10^5~M_\odot$ and $10^6~M_\odot$ systems, the time delay of gravitons are from a few hours to more than 10 days, which should be observed easily by the space-based detectors if $\lambda_{\rm_{g}} = 1.6 \times 10^{13}~\rm{km}$ . This show the potential of EMRI for improving the limit on $\lambda_{\rm_{g}}$.

\begin{table}[ph]
\tbl{The delay of arrival time of GW modes relative to the $k = 4$ mode (in seconds, $\lambda_{\rm_{g}}=1.6 \times 10^{13}~\rm{km}$). We set $D_L= 1.00~\rm{Gpc}$, where $Z \approx 0.20$ and $D \approx 0.83~\rm{Gpc}$. The parameters $e = 0.5$, $p = 12~M$, $a = 0.9$, the total masses are $M= 500~M_{\odot}$, $10^4~M_{\odot}$, $10^5~M_{\odot}$,  and $10^6~~M_{\odot}$ with the symmetric mass ratio $\nu= 10^{-2}$, $10^{-3}$, $10^{-4}$, and $10^{-5}$ respectively. }
{\begin{tabular}{ccccccccc} \toprule
&k = 2&k = 3&
 k= 4&k = 5&k = 6\\
\hline
$500~M_{\odot}$&$6.5845\times 10^{-1}$&$2.3996\times 10^{-1}$&$0$&$-1.5022\times 10^{-1}$&$-2.5046\times 10^{-1}$\\
$10^4~M_{\odot}$&$2.6341\times 10^{2}$&$9.6016\times 10^{1}$&$0$&$-6.0126\times 10^{1}$&$-1.0025\times 10^{2}$\\
$10^5~M_{\odot}$&$2.6328\times 10^{4}$&$9.5977\times 10^{3}$&$0$&$-6.0107\times 10^{3}$&$-1.0022\times 10^{4}$\\$10^6~M_{\odot}$&$2.6330\times 10^{6}$&$9.5982\times 10^{5}$&$0$&$-6.0110\times 10^{5}$&$-1.0023\times 10^{6}$\\ \botrule\end{tabular} \label{tab:table1}}
\end{table}

\begin{table}[ph]
\tbl{The delay of arrival time of GW modes relative to the $k = 10$ mode (in seconds, $\lambda_{\rm_{g}}=1.6 \times 10^{13}~\rm{km}$). We set $D_L= 1.00~\rm{Gpc}$, where $Z \approx 0.20$ and $D \approx 0.83~\rm{Gpc}$. The parameters $e = 0.7$, $p = 12~M$, $a = 0.9$, the total masses are $M= 500~M_{\odot}$, $10^4~M_{\odot}$, $10^5~M_{\odot}$,  and $10^6~~M_{\odot}$ with the symmetric mass ratio $\nu= 10^{-2}$, $10^{-3}$, $10^{-4}$, and $10^{-5}$ respectively.  }
{\begin{tabular}{ccccccccc} \toprule
&k = 8&k = 9&
 k= 10&k = 11&k = 12\\
\hline
$500~M_{\odot}$&$1.9265\times 10^{-1}$&$8.3765\times 10^{-2}$&$0$&$-6.5818\times 10^{-2}$&$-1.1847\times 10^{-1}$\\$10^4~M_{\odot}$&$7.7072\times 10^{1}$&$3.3514\times 10^{1}$&$0$&$-2.6336\times 10^{1}$&$-4.7406\times 10^{1}$\\$10^5~M_{\odot}$&$7.7100\times 10^{3}$&$3.3526\times 10^{3}$&$0$&$-2.6345\times 10^{3}$&$-4.7422\times 10^{3}$\\$10^6~M_{\odot}$&$7.7103\times 10^{5}$&$3.3527\times 10^{5}$&$0$&$-2.6345\times 10^{5}$&$-4.7423\times 10^{5}$\\\botrule\end{tabular} \label{tab:table2}}
\end{table}

\begin{figure}[!h]
\begin{center}
\includegraphics[height=1.5in]{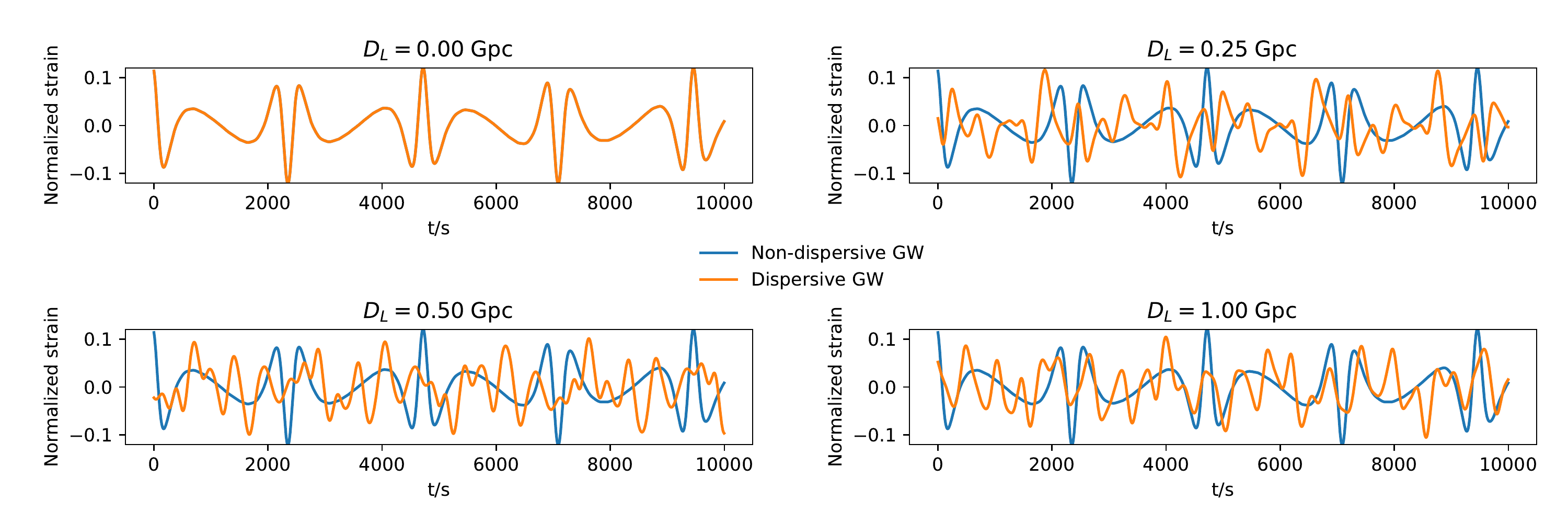}
\caption{The distortion of gravitational waves from an extreme-mass-ratio inspiral during propagation(EMRI,  $\lambda_{\rm_{g}} =1.6 \times 10^{13}~\rm{km}$). The total Mass $M = 10^6~M_{\odot}$, the symmetric mass ratio $\nu= 10^{-5}$, $e=0.5$, $p= 12 M$ and $a = 0.9$. We set the luminosity distance of sources $D_L$ from $0$ to $1.00$ Gpc. The orange curves represent the waveforms with dispersion during the propagation, and the cyan one is the nondispersive ones.
}  
\label{fig:distortion}
\end{center}
\end{figure}

As we mentioned before, for GWs with dispersion effect, the dephasing will appear compared to the one without dispersion. In other words, waveforms are distorted by the dispersion during propagation. Figure 1~\ref{fig:distortion} shows the distortion of GWs during propagation for  $M = 10^6~M_{\odot}$ case. The length of GW series is $10^4~\rm{s}$ without orbital evolution. The strain of GWs here are normalized, so that we could compare the waveforms of GWs in different $D_L$. It is clear that the waveform is distorted by dispersion effect during the propagation.

\begin{figure}[!h]
\begin{center}
\includegraphics[height=2.0in]{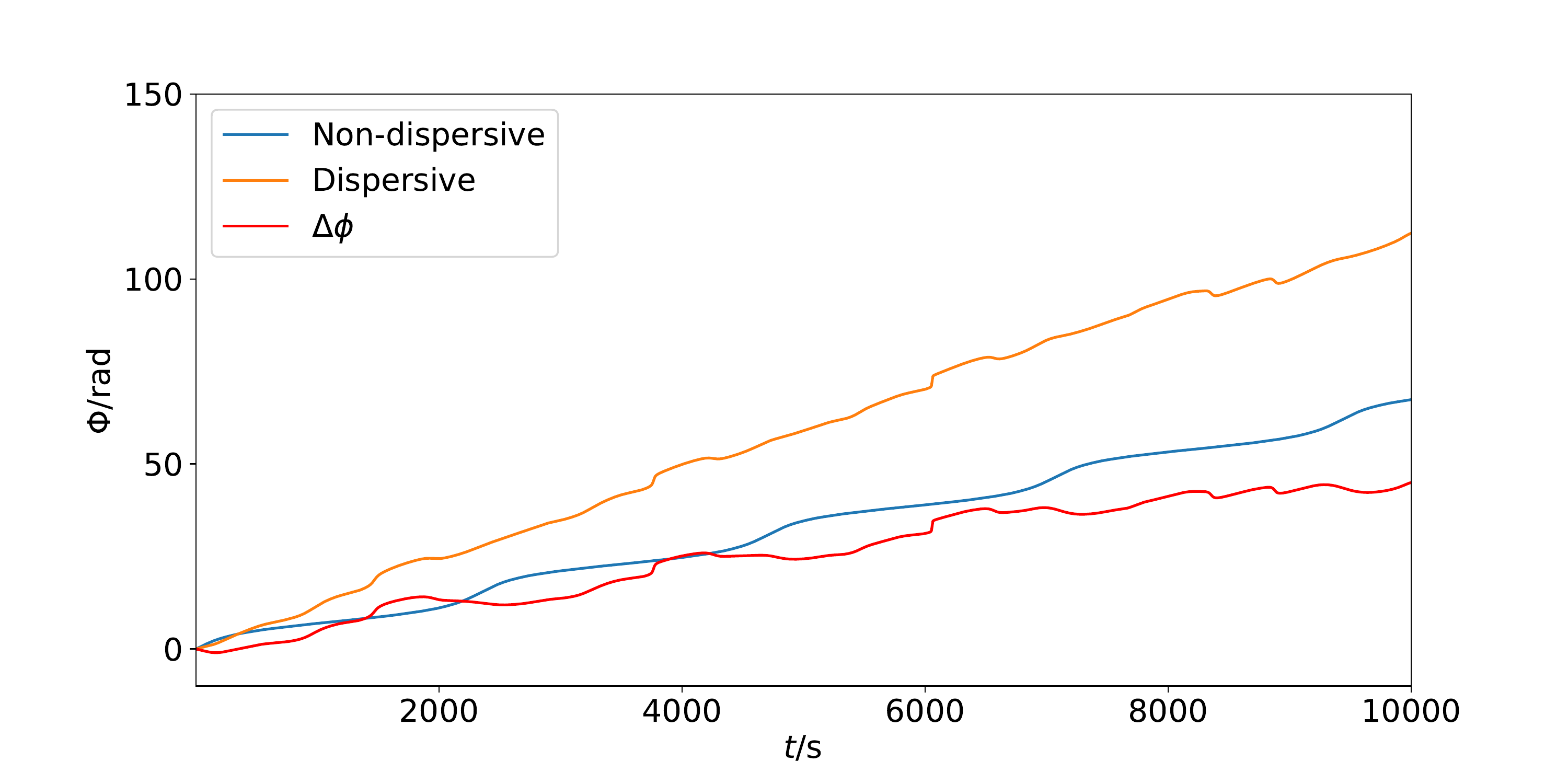}
\caption{The phase of gravitational waves in the dispersive case and nondispersive case and dephasing between two waveforms (EMRI,  $\lambda_{\rm_{g}} =1.6 \times 10^{13}~\rm{km}$). The total Mass $M = 10^6~M_{\odot}$, the symmetric mass ratio $\nu= 10^{-5}$, $e=0.5$, $p= 12 M$ and $a = 0.9$. We set $D_L= 1.00~\rm{Gpc}$, where $Z \approx 0.20$ and $D \approx 0.83~\rm{Gpc}$. The cyan, orange and red curves represent the nondispersive's, dispersion's phase and dephasing respectively.
}  
\label{fig:phase_difference}
\end{center}
\end{figure}

Obviously, the time delay will induce dephasing. In Fig.~\ref{fig:phase_difference}, by using aLIGO and AdV's lower limit \cite{abbott2017gw170104} on $\lambda_{\rm_{g}} = 1.6 \times 10^{13}~\rm{km}$, we plot the phase evolution of two waveforms with and without dispersion. It is shown that the dispersion effects can change the phase of GWs efficiently. The dephasing $\Delta \phi \left(= \phi_{\rm{Dispersive}} - \phi_{\rm{Nondispersive}}\right)$ is the phase difference between the dispersion waveform and the nondispersive one, and exceeds one radian in a very short duration(by requiring the Fourier phase dephasing be less than one radian, we could also get a constraint\cite{lindblom2008model} on $\lambda_{\rm_{g}}$). Considering LISA is very sensitive to the waveform phase, the difference of two waveforms should be recognized easily.

 For further discussion of the quantified difference between  dispersive GWs and nondispersive GWs, we adopt the well-known matched-filtering technology \cite{finn1992detection}. Given time series $h_1(t)$ and $h_2(t)$, the overlap of the two series is 
\begin{align}
{\rm Overlap} = \frac{\left<h_1, h_2\right>}{\sqrt{\left<h_1, h_1\right>\left<h_2, h_2\right>}},\label{eqn:ff}                          
\end{align} 
where $\left<h_1, h_2\right>$ is the symmetric inner product and it has the form
\begin{align}
\left<h_1, h_2\right> = \int^{\infty}_{-\infty}df\frac{\tilde{h_1}(f)\tilde{h_2^{\ast}}(f)}{S_n(|f|)},\label{eqn:ff}                          
\end{align} 
where the overhead tildes denotes the Fourier transform and the asterisk denotes complex conjugation. $S_n(|f|)$ is the spectral noise density curve, and in this work we use an analytic approximation \cite{babak2007kludge} to the  spectral noise density curve of LISA(from $10^{-4}~\rm{Hz}$ to $1.25 \times 10^{-1}~\rm{Hz}$) and   updated Advanced LIGO design curve\cite{barsotti2018updated}(from $10~\rm{Hz}$ to $8.192 \times 10^{3}~\rm{Hz}$). Another term is match, which is equivalant to overlap maximized over time and phase. The higher the match, the more similar the two time series are. For two same time series, the match is equal to $1$.  We use match to quantify the differences between the GW series with and without dispersion, and the results of match is calculated by PyCBC\cite{alex_nitz_2019_3247679}. 

\section{Results and analysis.}
\begin{figure}[!h]
\begin{center}
\includegraphics[height=2.0in]{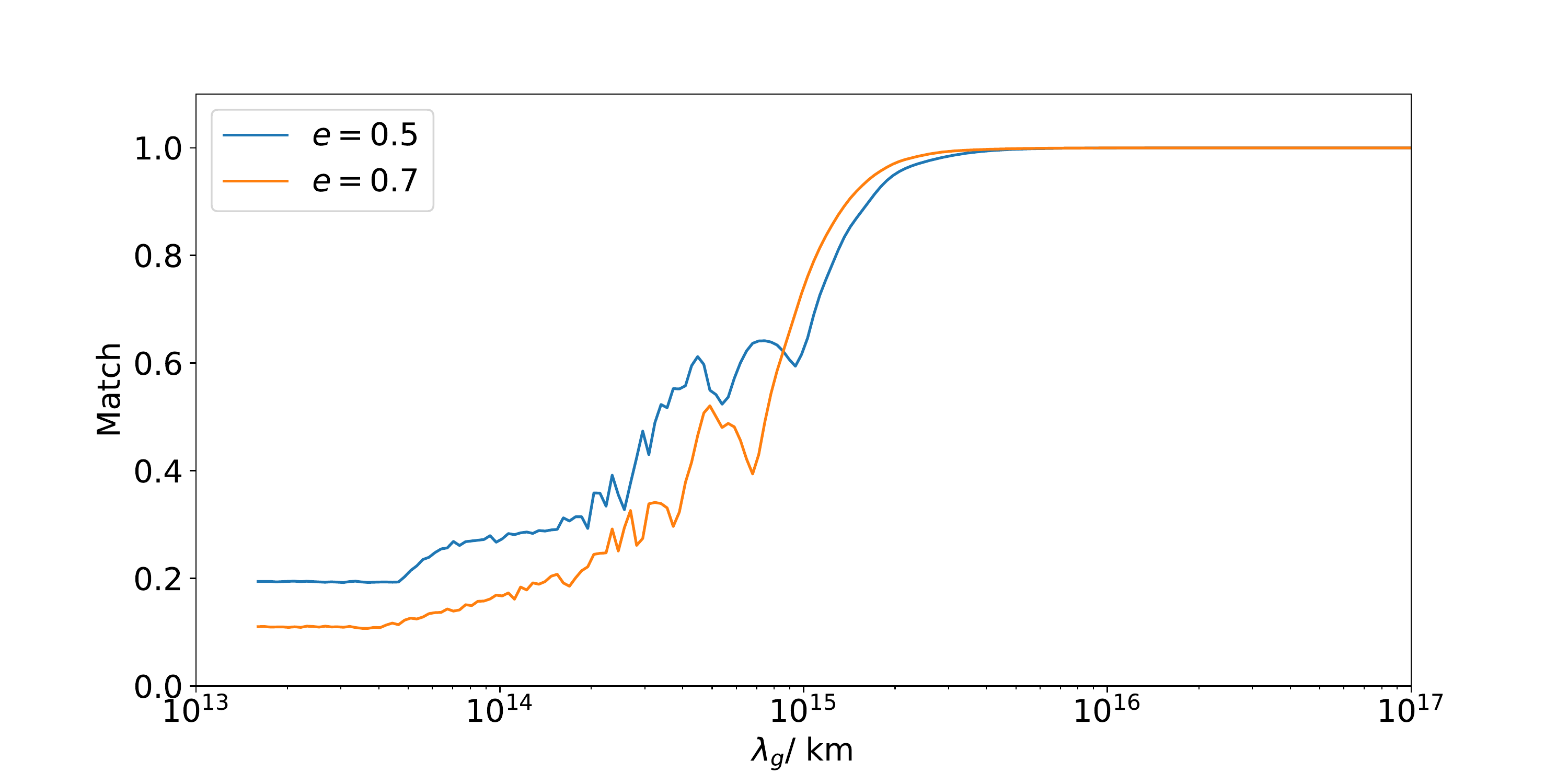}
\caption{Matches between dispersive and nondispersive GW series in different $e$ and $\lambda_{\rm_{g}}$ (LISA band). The curves begins at $\lambda_{g} = 1.6 \times 10^{13}~\rm{km}$, the total masses of systems $M=  10^6~M_{\odot}$, where the symmetric mass ratio $\nu= 10^{-5}$ , $e=0.5$ (cyan line) and $e=0.7$(orange line), $p= 12 M$ and $a = 0.9$. For GW series, the duration is $2^{25}~\rm{s}$ (about $1.06~\rm{year}$) and the sampling frequency is $0.25~\rm{Hz}$. We set $D_L= 1.00~\rm{Gpc}$, where $Z \approx 0.20$ and $D \approx 0.83~\rm{Gpc}$.}  
\label{fig:Fitting_factor_lambda_g0.5}
\end{center}
\end{figure}

\begin{figure}[!h]
\begin{center}
\includegraphics[height=2.0in]{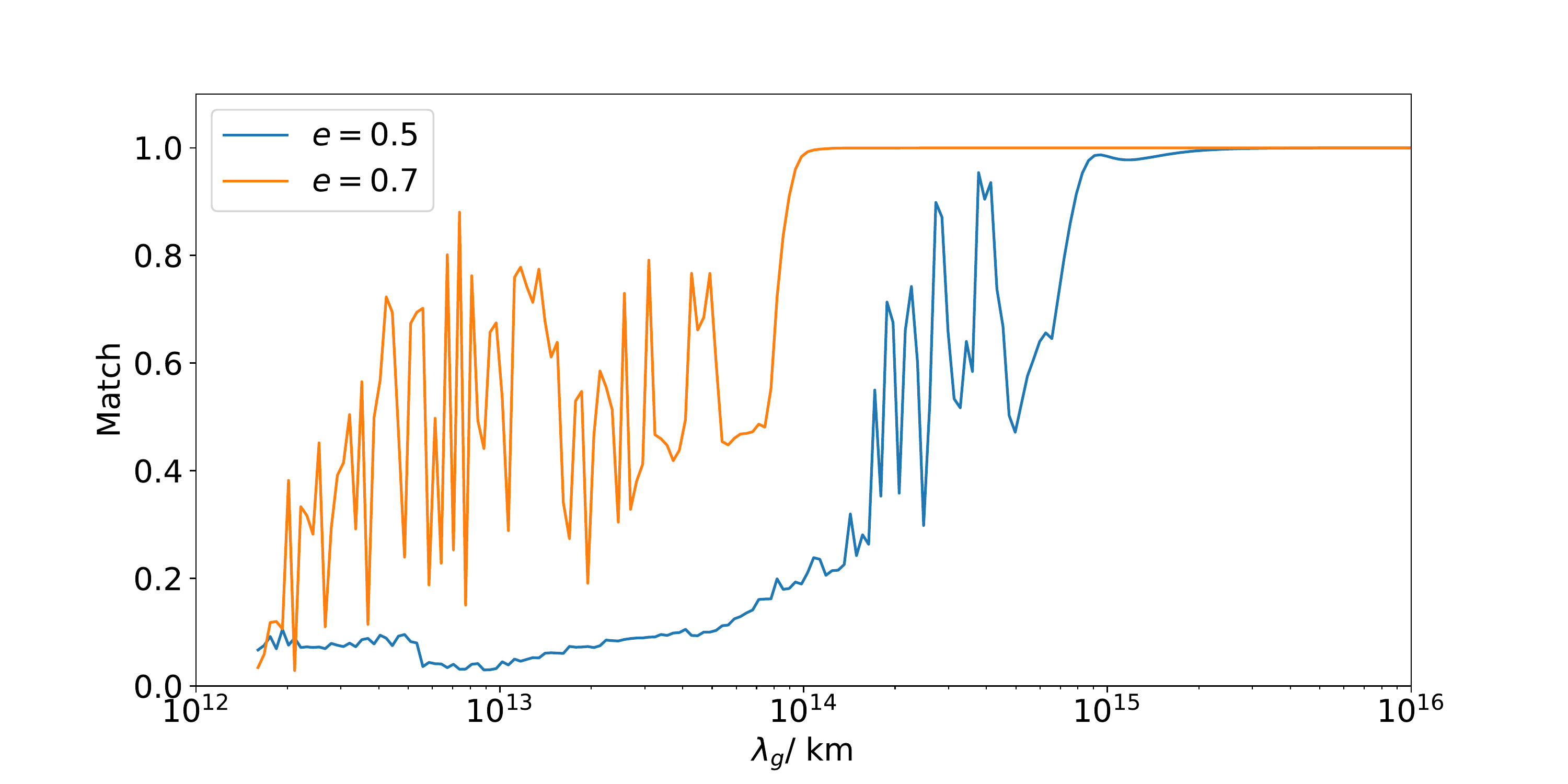}
\caption{Matches between dispersive and nondispersive GW series in different $e$ and $\lambda_{\rm_{g}}$ (aLIGO band). The curves begins at $\lambda_{g} = 1.6 \times 10^{12}~\rm{km}$, the total masses of systems $M= 500~M_{\odot}$, where the symmetric mass ratio $\nu= 10^{-2}$ , $e=0.5$ (cyan line) and $e=0.7$(orange line), $p= 12 M$ and $a = 0.95$. For GW series, the duration is $2^7~\rm{s}$ ($128~\rm{s}$) and the sampling frequency is $2^{14}~\rm{Hz}$.  We set $D_L= 1.00~\rm{Gpc}$, where $Z \approx 0.20$ and $D \approx 0.83~\rm{Gpc}$.}  
\label{fig:Fitting_factor_lambda_g0.7}
\end{center}
\end{figure}

We calculate the matches between the dispersive and nondispersive GW series from eccentric sources, by which we acquire the constraint on $\lambda_{\rm_{g}}$. All the GW series contain radiation reaction, and the waveform is calculated by Eq.~(\ref{eqn:Delta_t_a2}). For LISA cases, we provide the waveforms with the duration of  $2^{25}~\rm{s}$ (about $1.06~\rm{year}$, Fig.~\ref{fig:Fitting_factor_lambda_g0.5}) or $2^{16}~\rm{s}$(about $0.76~\rm{d}$, Fig.~\ref{fig:Fitting_factor_lambda_gD} and Fig.~\ref{fig:ff_e_p}), and the sampling frequency is $0.25~\rm{Hz}$. For aLIGO cases, we provide the waveforms with the duration of $2^{7}~\rm{s}$ ($=128 \rm{s}$, Fig.~\ref{fig:Fitting_factor_lambda_g0.7}), and sampling frequency is $2^{14}~\rm{Hz}$($16384 \rm{Hz}$). 

Figures ~\ref{fig:Fitting_factor_lambda_g0.5} and ~\ref{fig:Fitting_factor_lambda_g0.7} talk about the matches between dispersive and nondispersive GW series in different $e$ and $\lambda_{\rm_{g}}$. Then We could get the constraint on by setting a criteria\cite{lindblom2008model, chatziioannou2017constructing}  on $\lambda_{\rm_{g}}$. In this work, we set a criteria value 0.97 (This value is popular used in literature for example Ref.~45\comment{\cite{yunes2011extreme}}). If the match value is less than 0.97, then we say that we could distinguish the dispersive and nondispersive GW series. The corresponding value of $\lambda_{\rm_{g}}$ when $\mathrm{match} = 0.97$ could be regarded as a lower limit of $\lambda_{\rm_{g}}$. In Fig.~\ref{fig:Fitting_factor_lambda_g0.5}, for $e = 0.5$ and $e = 0.7$ , we could get lower limits of  $\lambda_{\rm_{g}} \approx 2.4 \times 10^{15}~\rm{km} ~\rm{and}~ 2.0 \times 10^{15}~\rm{km}$. In Fig.~\ref{fig:Fitting_factor_lambda_g0.7}, we could get lower limits of  $\lambda_{\rm_{g}} \approx 8.7 \times 10^{14}~\rm{km} ~\rm{and}~ 9.8 \times 10^{13}~\rm{km}$. All the results above give a more strigent constraint on $\lambda_{\rm{g}}$ than the current result from LIGO observations \cite{abbott2017gw170104}, which show the protential of EMRIs/IMRIs on the constraint of $\lambda_{\rm_{g}}$. Meanwhile, it seems ground-based detectors such as aLIGO also could test this.

\begin{figure}[!h]
\begin{center}
\includegraphics[height=2.0in]{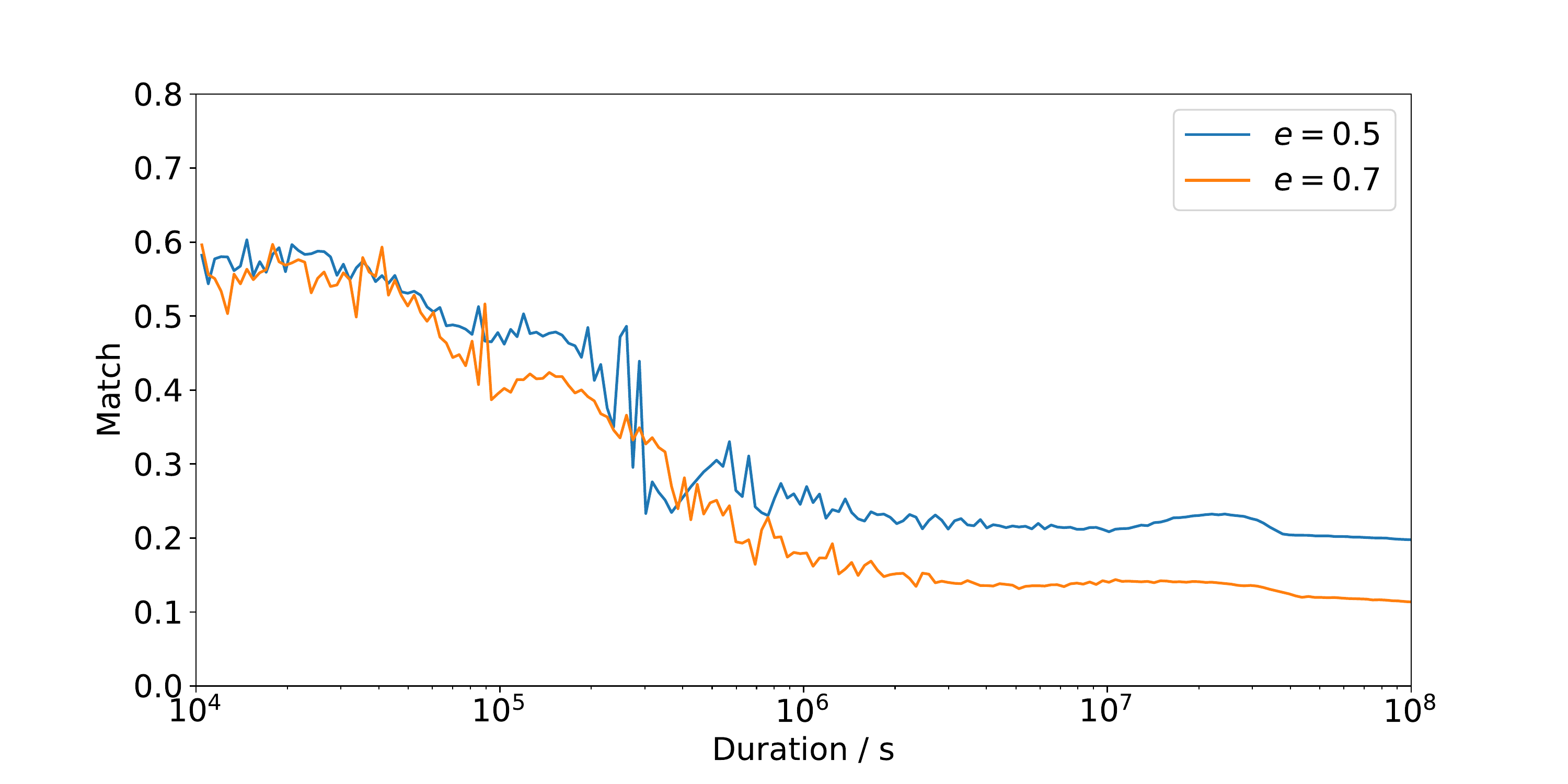}
\caption{Matches between dispersive and nondispersive GW series in different duration(LISA band).The total masses of systems $M=  10^6~M_{\odot}$, where the symmetric mass ratio $\nu= 10^{-5}$ , $e=0.5$ (cyan line) and $e=0.7$(orange line), $p= 12 M$ and $a = 0.9$. For GW series, the duration is $2^{25}~\rm{s}$ (about $1.06~\rm{year}$) and the sampling frequency is $0.25~\rm{Hz}$. We set $D_L= 1.00~\rm{Gpc}$, where $Z \approx 0.20$ and $D \approx 0.83~\rm{Gpc}$.}  
\label{fig:Overlap_1e6M}
\end{center}
\end{figure}

\begin{figure}[!h]
\begin{center}
\includegraphics[height=2.0in]{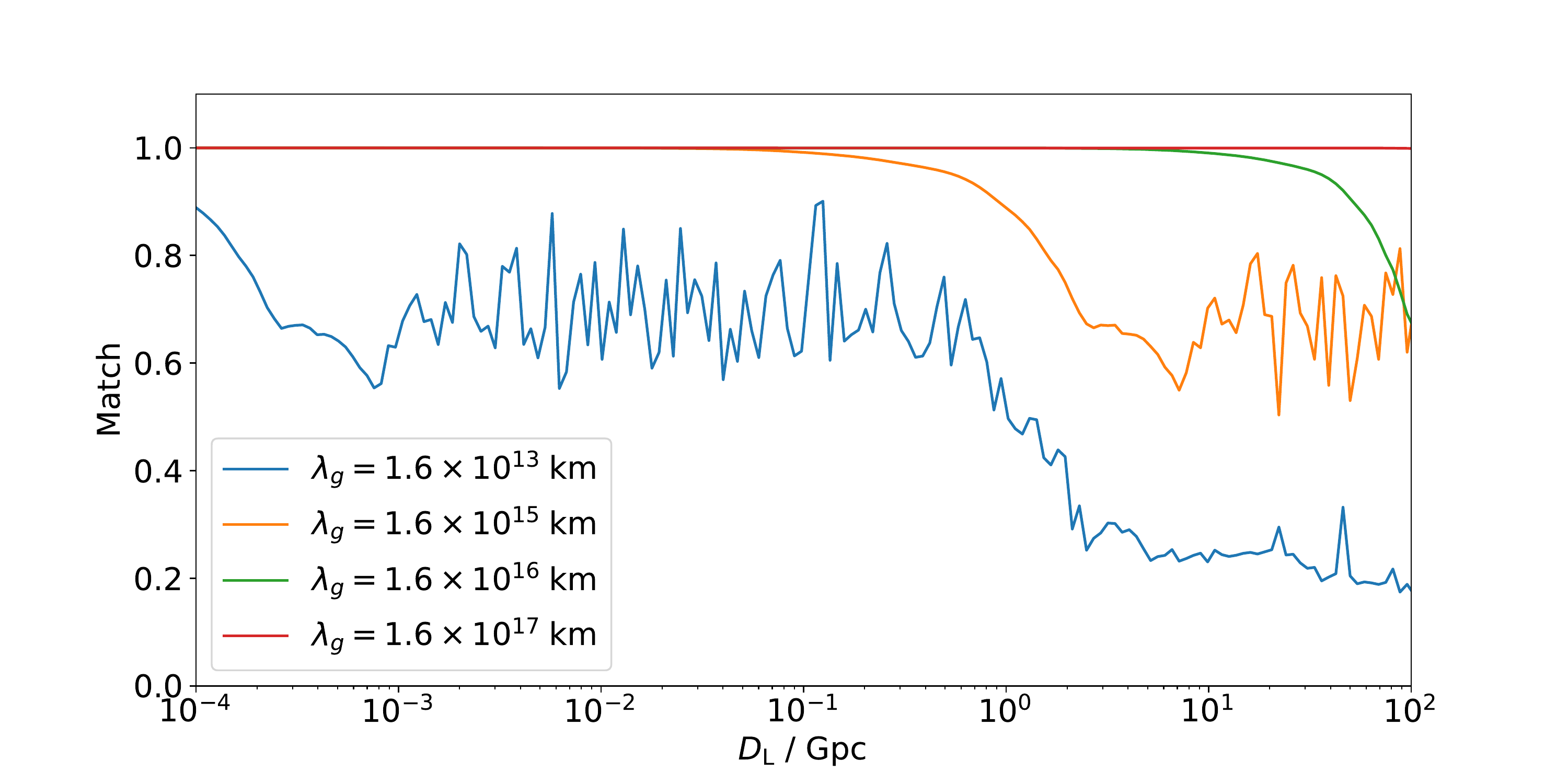}
\caption{Matches between dispersive and nondispersive GW series in different  and $D_L$. We set the Compton wavelength of the graviton $\lambda_{\rm_{g}} = 1.6 \times 10^{13}$ km(cyan line), $1.6 \times 10^{15}$ km(orange line), $1.6 \times 10^{16}$ km(green line) and $1.6 \times 10^{17}$ km(red line). The total masses of systems $M=10^6~M_{\odot}$, the symmetric mass ratio $\nu=10^{-5}$, $e=0.5$ and $p= 12 M$.  We set $D_L= 1.00~\rm{Gpc}$, where $Z \approx 0.20$ and $D \approx 0.83~\rm{Gpc}$.}  
\label{fig:Fitting_factor_lambda_gD}
\end{center}
\end{figure}

\begin{figure}[!h]
\begin{center}
\includegraphics[height=3.5in]{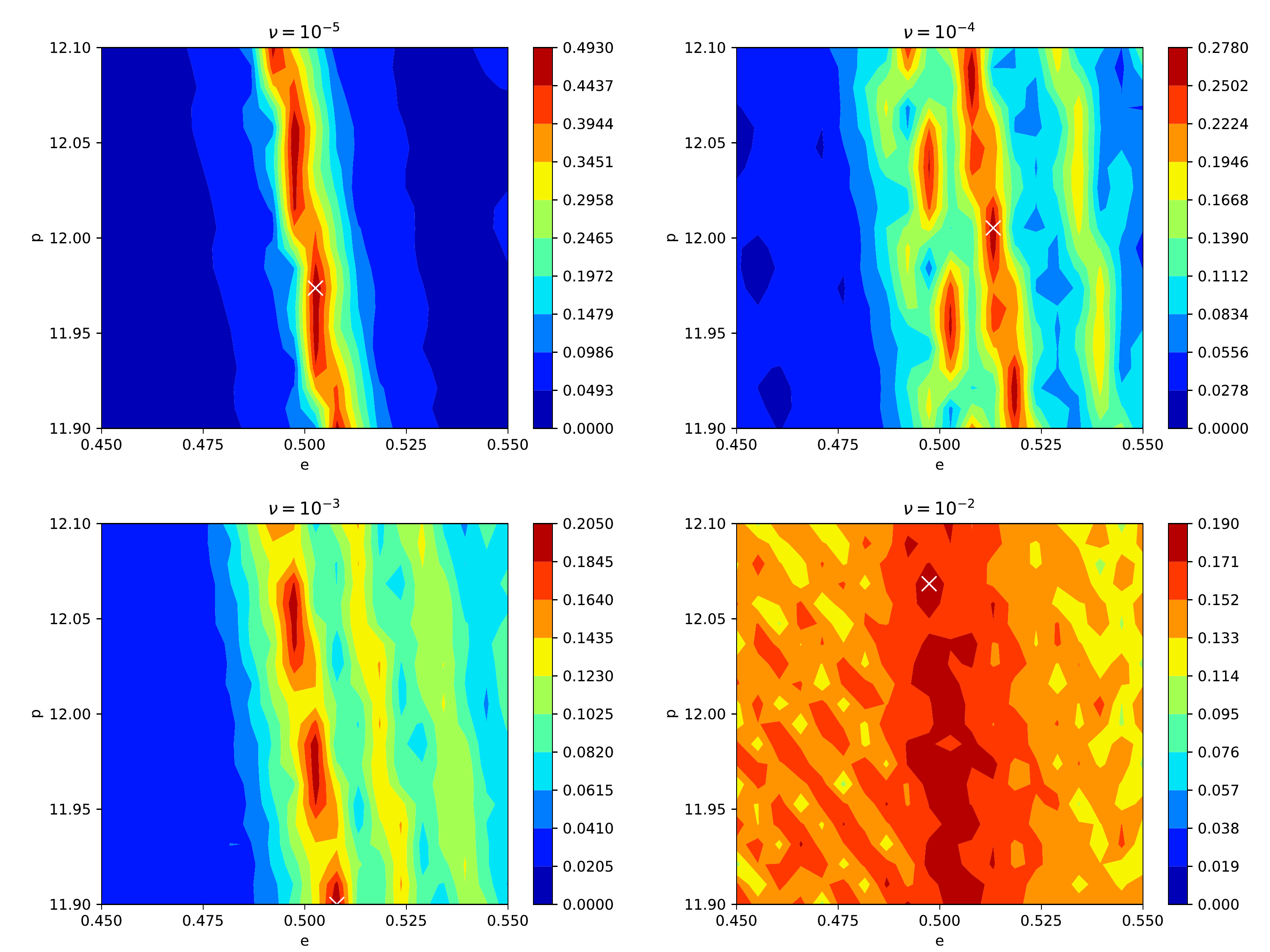}

\caption{Contour of matches between dispersive GW series and nondispersive GW series in different $e$ and $p$(LISA band, and $\lambda_{\rm_{g}} = 1.6 \times 10^{13}~\rm{km}$). The dispersive GW series is in $e=0.5$ and $p= 12 M$. The white cross show the largest match value. Other parameters are the same, the total masses of systems $M = 10^6~M_{\odot}$, and the symmetric mass ratio $\nu= 10^{-5}$, $10^{-4}$, $10^{-3}$ and $10^{-2}$ respectively, and $a = 0.9$. We set the luminosity distance $D_L = 1.00~\rm{Gpc}$, where $Z \approx 0.20$ and $D \approx 0.83~\rm{Gpc}$}  \label{fig:ff_e_p}
\end{center}
\end{figure}

 For LISA case, as illustrated in Fig.~\ref{fig:Fitting_factor_lambda_g0.5}, the matches increase roughly as the $\lambda_{\rm_{g}}$ goes up. This is easy to understand based on Eq.~(\ref{eqn:Delta_t_a2}). A larger $\lambda_{\rm_{g}}$ usually means a lower time delay between two modes. We could find $e = 0.5$ case is close to $e = 0.7$ case. This is because for the $M = 10^6 ~M_{\odot} (\nu= 10^{-5})$ system, the orbital evolution is slow, even with a duration of $1.06 \rm~{year}$. For aLIGO case in Fig.~\ref{fig:Fitting_factor_lambda_g0.7},  we could notice that the $e = 0.5$ case is better than  $e = 0.7$ case on the constraint on $\lambda_{\rm{g}}$. This is because the $M = 500 ~M_{\odot} (\nu= 10^{-2})$ system has a swift orbit evolution, and $e = 0.5$ case went through a rapider change in $e$ and $p$ than  $e = 0.7$ case in $128\rm~{s}$. According to our calculation, at the end $e = 0.5$ case is next to the innermost stable circular orbit (ISCO). Therefore  $e = 0.5$ case got more accumulated phase differences, which lead to more strigent constraint on $\lambda_{\rm{g}}$. In Fig.~\ref{fig:Overlap_1e6M}, we calculate the matches between dispersive and nondispersive GW series in different duration. We could find that the longer series have lower matches roughtly, which reflect the accumulated phase differences. 
 
The constraint capability on $\lambda_{\rm{g}}$ also depends on the distance of GW source based on Eq.~(\ref{eqn:Delta_t_a2}). The variation of matches with luminosity distance $D_L$ are shown in Fig.~\ref{fig:Fitting_factor_lambda_gD}.  We use the GW series with the duration of $2^{16}~\rm{s}$ (about $0.76~\rm{day}$) to show the basic idea. In Eq.~(\ref{eqn:Delta_t_a2}), we could see for all cases except $\lambda_{\rm_{g}} = 1.6 \times 10^{17}~\rm{km}$ case, the matches decrease with the increase of $D_L$. We can see that if we expect a two-order improvement on the constraint of $\lambda_{\rm g}$,  an EMRI at distance larger than 0.1 Gpc is needed(this is for the $2^{16}~\rm{s}$  GW series, in fact we should use longer series so the distance would be less than 0.1 Gpc). Fortunately, this distance is typical for EMRIs and can be easily detected by LISA.

Though we demonstrate the obvious mismatch between two waveforms with and without dispersion, one may doubt that if the dispersive GW series would be matched by a nondispersive mimic through adjusting orbital parameters or not. In other words, if the answer is yes, one could not distinguish the dispersive GW signals because of the huge parameter space of nondispersive waveform templates. This is the so-called ``confusion" problem. While in our case, the variation of $M$ and $a$ ($p$, $e$ also) result in the change of the fundamental frequency of GW, which influence all the frequencies of GW modes. It is not likely that two GW series with different frequencies of modes would have a high match. As a demonstration, we calculate the matches between several dispersive GW series and numerous nondispersive GW templates.  Figure~\ref{fig:ff_e_p} shows the matches between some certain dispersive GW series and nondispersive GW templates by scanning the parameters $e$ and $p$. For each dispersive waveform, we use 400 nondispersive waveforms to match. As illustrated, for all cases (typical EMRIs or IMRIs), the highest matches  are less than $0.5$, and no exception is found yet in our other calculations. It means that the confusion problem does not exist.  Consequently, for the GWs with a large enough dispersion effect, we speculate that we could catch the gravitational dispersion with appropriate GW templates.

\section{Conclusion}
To conclude, in this paper we demonstrate the great potential of EMRI/IMRI signals on the test of GW dispersion (constraint on $\lambda_{\rm{g}}$) , which should be an important scientific target for space-based detectors. We impose the dispersion effect to the gravitational waveforms of EMRIs/IMRIs for which the orbital eccentricities could still be large at the final stage of evolution. The eccentricities lead to varied GW modes to be emitted at the same moment, but arrive at different moments if dispersion really exists. We demonstrate that the GW dispersion of eccentric sources will cause obvious dephasing and distortion of the waveforms, which could be observed by space-based detectors. Our results show that the observations of EMRIs/IMRIs with LISA will constrain the $\lambda_{\rm_{g}}$ two or three orders better than the current level from LIGO observations. Just for attention, Jones \cite{jones2004bounding} used scaling relations of the dephasing of waveforms rather than the matched filtering, he find three orders of magnitude stronger for EMRIs than ours. However, we now do not find the exact reason of this difference. In addition, we also investigate this dispersion effect for eccentric IMRIs in LIGO band, and fortunately aLIGO may able to detect this effect if an IMRI could be detected, and will constrain the $\lambda_{\rm{g}}$ better than the current level.

EMRIs with circular orbits can also be used to test dispersion if adopting the same method LIGO used \cite{abbott2017gw170104}. Considering that the frequency evolution is very slow due to the extremely small mass-ratio, one need a long observation to do this. Therefore there are some challenges and chances. On the one hand, accurate templates with long evolution are necessary, and some people are working on this\cite{van2018fast}. On the other hand, during one-year orbital evolution, unsteady and nongaussian noises from both GW source and detector could happen, and produce uncertainty on the parameter estimation. In general, with efforts on the future space-based detectors, the EMRIs/IMRIs, eccentric or circular, should be a potential tool for testing gravity theory. 

\section*{Acknowledgments}
This work is supported by NSFC No. 11773059, and we also appreciate the anonymous Referee’s suggestions about our work. This work was also supported by MEXT, JSPS Leading-edge Research Infrastructure Program, JSPS Grant-in-Aid for Specially Promoted Research 26000005, JSPS Grant-in-Aid for Scientific Research on Innovative Areas 2905: JP17H06358, JP17H06361 and JP17H06364, JSPS Core-to-Core Program A. Advanced Research Networks, JSPS Grant-in-Aid for Scientific Research (S) 17H06133, the joint research program of the Institute for Cosmic Ray Research, University of Tokyo, and by Key Research Program of Frontier Sciences, CAS, No. QYZDB-SSW-SYS016.


\appendix

\section{Teukolsky equation}

The gravitational perturbation of Kerr space-time is described by the Teukolsky equation by the Weyl curvature (complex) scalar $\psi_4$, decomposed in frequency domain $\psi_4=\rho^4\int^{+\infty}_{-\infty}{d\omega\sum_{lm}{R_{lm\omega}(r)_{~-2}S^{a\omega}_{lm}(\theta)e^{im\phi}e^{-i\omega
			t}}}$ with spin-weighted spheroidal harmonics $_{~-2}S^{a\omega}_{lm}$, obeys \cite{teukolsky1973perturbations}:

\begin{equation}
\Delta^2\frac{d}{dr}\left(\frac{1}{\Delta}\frac{d
	R_{lm\omega}}{dr}\right)-V(r)R_{lm\omega}=-\mathcal{T}_{lm\omega}(r),
\label{Teukolsky}
\end{equation}
where $\mathcal{T}_{lm\omega}(r)$ is the source term, which is connected by the stress-energy tensor of the perturbation source, and the potential is
\begin{equation}
V(r)=-\frac{K^2+4i(r-M)K}{\Delta}+8i\omega r+\lambda,
\end{equation}
where $K=(r^2+a^2)\omega-ma, ~\lambda=E_{lm}+a^2\omega^2-2a m w-2$ and $\Delta = r^2-2Mr+a^2$. 

First, we consider the homogeneous Teukolsky equation where the source term is zero. We can solve it by analytical expansion, as discussed in \cite{sasaki2003analytic,fujita2004new} and here we don't go into technical details of it. The homogeneous Teukolsky equation allows two independent solutions $R^{\rm H}_{lm\omega}$, which is purely ingoing at the horizon, and $R^\infty_{lm\omega}$, which is purely outgoing at infinity:
\begin{align}\nonumber
\label{RH}
R^{\rm H}_{lm\omega}&=B^{\rm hole}_{lm\omega}\Delta^2 e^{-ipr*},\quad
r\rightarrow r_+\\
R^{\rm H}_{lm\omega}&=B^{out}_{lm\omega}r^3 e^{i\omega
	r*}+r^{-1}B^{\rm in}_{lm\omega} e^{-i\omega r*},\quad r\rightarrow
\infty;
\end{align}
\begin{align}\nonumber
\label{Rinf}
R^{\infty}_{lm\omega}&=D^{\rm{out}}_{lm\omega} e^{ip r*}+\Delta^2
D^{\rm in}_{lm\omega} e^{-ip r*},\quad
r\rightarrow r_+\\
R^{\infty}_{lm\omega}&=r^3 D^{\infty}_{lm\omega} e^{i\omega
	r*},\quad r\rightarrow \infty,
\end{align}
where $p= \omega-\frac{ma}{2Mr_+}$, $r_+=M+\sqrt{M^2-a^2}$ and $r*$ is the tortoise coordinate related to $r$ by $dr*/dr = (r^2+a^2)/\Delta$

Then, using the homogeneous solutions and proper boundary conditions, we can construct the solution to radial Teukolsky equation with source term. By imposing BH boundary condition, i.e. wave being purely outgoing at infinity and purely ingoing at horizon, the radial function is:
\begin{align}
\label{BHsolution}
	R^{\rm BH}_{lm\omega}(r)=\frac{R^{\infty}_{lm\omega}(r)}{2i\omega
		B^{\rm in}_{lm\omega}D^{\infty}_{lm\omega}}\int^{r}_{r_+}{dr'\frac{R^{\rm H}_{lm\omega}(r')\mathcal{T}_{lm\omega}(r')}{\Delta(r')^2}}+ \\ \frac{R^{\rm H}_{lm\omega}(r)}{2i\omega
		B^{\rm in}_{lm\omega}D^{\infty}_{lm\omega}}\int^{\infty}_{r}{dr'\frac{R^\infty_{lm\omega}(r')\mathcal{T}_{lm\omega}(r')}{\Delta(r')^2}}
\end{align}

The asymptotic behavior of this solution near horizon and infinity is:
\begin{align}
\label{eq_RBH_asym}
R^{\rm BH}_{lm\omega}(r\rightarrow \infty)&=Z^{\rm H}_{lm\omega}r^3 e^{i\omega
	r*},\\
R^{\rm BH}_{lm\omega}(r\rightarrow r_+)&=Z^{\infty}_{lm\omega}\Delta^2 e^{-ip
	r*}.
\end{align}

By taking the limit at $r\rightarrow \infty$ and $r\rightarrow r_+$ of the solution (Eq. \ref{BHsolution}), with the asymptotic behavior of homogeneous solutions (Eq. \ref{RH}, \ref{Rinf}), one can find the amplitudes $Z^{H,\infty}_{lm\omega}$:
\begin{equation}
	Z^{\rm H}_{lm\omega} = \frac{1}{2i\omega B^{\rm in}_{lm\omega}} \int_{r_+}^{r} dr' \frac{R^{\rm H}_{lm\omega}(r') \mathcal T _{lm\omega}(r') }{\Delta(r')^2 }
\end{equation}
\begin{equation}
Z^\infty_{lm\omega} = \frac{B^{\rm H}_{lm\omega}}{2i\omega B^{\rm in}_{lm\omega} D^{\infty }_{lm\omega}} \int_{r}^{\infty} dr' \frac{R^\infty_{lm\omega}(r') \mathcal T _{lm\omega}(r') }{\Delta(r')^2 }
\end{equation}
%
%


\begin{thebibliography}{10}

\bibitem{abbott2016observation}
{LIGO Scientific and Virgo
  Collabs. (B. P. Abbott {\it et al.)}}, {\em Phys. Rev. Lett.} {\bf 116}  (2016)   061102.

\bibitem{abbott2016gw151226}
{LIGO Scientific and Virgo
  Collabs. (B. P. Abbott {\it et al.)}}, {\em Phys. Rev. Lett.} {\bf 116}  (2016)   241103.

\bibitem{abbott2017gw170104}
{LIGO Scientific and Virgo
  Collabs. (B. P. Abbott {\it et al.)}}, {\em Phys. Rev. Lett.} {\bf 118}  (2017)   221101.

\bibitem{abbott2017gw170608}
{LIGO Scientific and Virgo
  Collabs. (B. P. Abbott {\it et al.)}}, {\em Astrophys. J. Lett.} {\bf 851}  (2017)   L35.

\bibitem{abbott2017gw170814}
{LIGO Scientific and Virgo
  Collabs. (B. P. Abbott {\it et al.)}}, {\em Phys. Rev. Lett.} {\bf 119}  (2017)   141101.

\bibitem{abbott2017gw170817}
{LIGO Scientific and Virgo
  Collabs. (B. P. Abbott {\it et al.)}}, {\em Phys. Rev. Lett.} {\bf 119}  (2017)   161101.

\bibitem{amelino2010doubly}
G.~Amelino-Camelia, {\em SIGMA} {\bf 2}  (2010) 230.

\bibitem{sefiedgar2011modified}
A.~S. Sefiedgar, K.~Nozari and H.~R. Sepangi, {\em Phys. Lett. B} {\bf 696}
  (2011) 119.

\bibitem{hovrava2009quantum}
P.~Ho{\v{r}}ava, {\em Phys. Rev. D} {\bf 79}  (2009)   084008.

\bibitem{garattini2011modified}
R.~Garattini and G.~Mandanici, {\em Phys. Rev. D} {\bf 83}  (2011)   084021.

\bibitem{jones2004bounding}
D.~Jones, {\em Astrophys. J. Lett.} {\bf 618}  (2004)   L115.

\bibitem{amaro2007intermediate}
P.~Amaro-Seoane, J.~R. Gair, M.~Freitag, M.~C. Miller, I.~Mandel, C.~J. Cutler
  and S.~Babak, {\em Class. Quantum Grav.} {\bf 24}  (2007)   R113.

\bibitem{danzmann1996lisa}
K.~Danzmann, L.~S. Team {\em et~al.}, {\em Class. Quantum Grav.} {\bf 13}
  (1996)   A247.

\bibitem{hu2017the}
W.-R. Hu and Y.-L. Wu, {\em Natl. Sci. Rev.} {\bf 4}  (2017) 685.

\bibitem{luo2016tianqin}
J.~Luo, L.-S. Chen, H.-Z. Duan, Y.-G. Gong, S.~Hu, J.~Ji, Q.~Liu, J.~Mei,
  V.~Milyukov, M.~Sazhin {\em et~al.}, {\em Class. Quantum Grav.} {\bf 33}
  (2016)   035010.

\bibitem{mirshekari2012constraining}
S.~Mirshekari, N.~Yunes and C.~M. Will, {\em Phys. Rev. D} {\bf 85}  (2012)
  024041.

\bibitem{Zhang2018Cosmological}
Y.~Zhang, X.~Liu, J.~Qi and H.~Zhang, {\em J. Cosmol. Astropart. Phys.} {\bf
  2018} (August 2018) 027.

\bibitem{Kostelecky2016testing}
V.~A. Kosteleck{\'{y}} and M.~Mewes, {\em Phys. Lett. B} {\bf 757} (June 2016)
  510.

\bibitem{abbott2017Multi-messenger}
{B. P. Abbott {\it et al.}}, {\em Astrophys. J. Lett.} {\bf 848}  (2017)   L12.

\bibitem{abbott2017gravitational}
{B. P. Abbott {\it et al.}}, {\em Astrophys. J. Lett.} {\bf 848}  (2017)   L13.

\bibitem{Fan2017speed}
X.-L. Fan, K.~Liao, M.~Biesiada, A.~Pi{\'o}rkowska-Kurpas and Z.-H. Zhu, {\em
  Phys. Rev. Lett.} {\bf 118} (March 2017) 688.

\bibitem{will1998bounding}
C.~M. Will, {\em Phys. Rev. D} {\bf 57}  (1998)   2061.

\bibitem{riess1998observational}
A.~G. Riess, A.~V. Filippenko, P.~Challis, A.~Clocchiatti, A.~Diercks, P.~M.
  Garnavich, R.~L. Gilliland, C.~J. Hogan, S.~Jha, R.~P. Kirshner {\em et~al.},
  {\em Astron. J.} {\bf 116}  (1998)   1009.

\bibitem{peters1963gravitational}
P.~Peters and J.~Mathews, {\em Phys. Rev.} {\bf 131}  (1963)   435.

\bibitem{han2017excitation}
W.-B. Han, Z.~Cao and Y.-M. Hu, {\em Class. Quantum Grav.} {\bf 34}  (2017)
  225010.

\bibitem{buonanno1999effective}
A.~Buonanno and T.~Damour, {\em Phys. Rev. D} {\bf 59}  (1999)   084006.

\bibitem{han2014gravitational}
W.-B. Han, {\em Int. J. Mod. Phys. D} {\bf 23}  (2014)   1450064.

\bibitem{cao2017waveform}
Z.~Cao and W.-B. Han, {\em Phys. Rev. D} {\bf 96}  (2017)   044028.

\bibitem{sago2015calculation}
N.~Sago and R.~Fujita, {\em Prog. Theor. Exp. Phys.} {\bf 2015}  (2015)
  073E03.

\bibitem{moreno1995gravitational}
C.~Moreno-Garrido, E.~Mediavilla and J.~Buitrago, {\em Mon. Notices Royal
  Astron. Soc} {\bf 274}  (1995) 115.

\bibitem{barack2004lisa}
L.~Barack and C.~Cutler, {\em Phys. Rev. D} {\bf 69}  (2004)   082005.

\bibitem{pierro2001fast}
V.~Pierro, I.~Pinto, A.~Spallicci, E.~Laserra and F.~Recano, {\em Mon. Notices
  Royal Astron. Soc} {\bf 325}  (2001) 358.

\bibitem{yunes2009post}
N.~Yunes, K.~Arun, E.~Berti and C.~M. Will, {\em Phys. Rev. D} {\bf 80}  (2009)
    084001.

\bibitem{teukolsky1973perturbations}
S.~A. Teukolsky, {\em Astrophys. J.} {\bf 185}  (1973)   635.

\bibitem{han2010gravitational}
W.-B. Han, {\em Phys. Rev. D} {\bf 82}  (2010)   084013.

\bibitem{han2011constructing}
W.-B. Han and Z.~Cao, {\em Physical Review D} {\bf 84}  (2011)   044014.

\bibitem{fujita2009efficient}
R.~Fujita, W.~Hikida and H.~Tagoshi, {\em Prog. Theor. Phys.} {\bf 121}  (2009)
  843.

\bibitem{hinshaw2013nine}
G.~Hinshaw, D.~Larson, E.~Komatsu, D.~Spergel, C.~Bennett, J.~Dunkley,
  M.~Nolta, M.~Halpern, R.~Hill, N.~Odegard {\em et~al.}, {\em Astrophys. J.
  Suppl. S.} {\bf 208}  (2013)  ~19.

\bibitem{lindblom2008model}
L.~Lindblom, B.~J. Owen and D.~A. Brown, {\em Phys. Rev. D} {\bf 78}  (2008)
  124020.

\bibitem{finn1992detection}
L.~S. Finn, {\em Phys. Rev. D} {\bf 46}  (1992)   5236.

\bibitem{babak2007kludge}
S.~Babak, H.~Fang, J.~R. Gair, K.~Glampedakis and S.~A. Hughes, {\em Phys. Rev.
  D} {\bf 75}  (2007)   024005.

\bibitem{barsotti2018updated}
{L. Barsotti}, {S. Gras}, {M. Evans} and {P. Fritschel}, {\em LIGO Document
  T1800044-v5,}   (2018) Updated Advanced LIGO sensitivity design curve
  (https://dcc.ligo.org/LIGO-T1800044/public).

\bibitem{alex_nitz_2019_3247679}
A.~Nitz, I.~Harry, D.~Brown, C.~M. Biwer, J.~Willis, T.~D. Canton, C.~Capano,
  L.~Pekowsky, T.~Dent, A.~R. Williamson, S.~De, M.~Cabero, B.~Machenschalk,
  D.~Macleod, P.~Kumar, S.~Reyes, T.~Massinger, G.~Davies, M.~Tapai, dfinstad,
  S.~Fairhurst, S.~Khan, A.~Nielsen, shasvath, F.~Pannarale, L.~Singer,
  H.~Gabbard, idorrington92, L.~M. Zertuche and B.~U.~V. Gadre, gwastro/pycbc:
  Pycbc release v1.14.0 (June 2019).

\bibitem{chatziioannou2017constructing}
K.~Chatziioannou, A.~Klein, N.~Yunes and N.~Cornish, {\em Phys. Rev. D} {\bf
  95}  (2017)   104004.

\bibitem{yunes2011extreme}
N.~Yunes, A.~Buonanno, S.~A. Hughes, Y.~Pan, E.~Barausse, M.~C. Miller and
  W.~Throwe, {\em Phys. Rev. D} {\bf 83}  (2011)   044044.

\bibitem{van2018fast}
M.~Van De~Meent and N.~Warburton, {\em Class. Quantum Grav} {\bf 35}  (2018)
  144003.

\bibitem{sasaki2003analytic}
M.~Sasaki and H.~Tagoshi, {\em Living Rev. Relativ.} {\bf 6}  (2003)  ~6.

\bibitem{fujita2004new}
R.~Fujita and H.~Tagoshi, {\em Prog. Theor. Phys.} {\bf 112}  (2004) 415.

\end{thebibliography}

\end{document}